\documentclass[preprint]{aastex}

\usepackage{color}
\usepackage{lineno}
\usepackage{rotating}
\usepackage{lscape}

\newcommand{\pks}{PKS\,0625$-$354}
\newcommand{\tc}{3C\,78}

\shorttitle{Suzaku Observations of $\gamma$-Ray Bright FR-I Radio Galaxies}
\shortauthors{Y. Fukazawa et al.}

\begin{document}

\title{{\em Suzaku} Observations of $\gamma$-Ray Bright Radio Galaxies:\\ Origin of the X-ray Emission and Broad-Band Modeling}

\author{Yasushi Fukazawa\altaffilmark{1,2}, Justin Finke\altaffilmark{3}, \L ukasz Stawarz\altaffilmark{4,5}, Yasuyuki Tanaka\altaffilmark{2}, Ryosuke Itoh\altaffilmark{1}, and Shin'ya Tokuda\altaffilmark{1}}

\email{\texttt{fukazawa@hep01.hepl.hiroshima-u.ac.jp}}

\altaffiltext{1}{Department of Physical Science, Hiroshima University, 1-3-1 Kagamiyama, Higashi-Hiroshima, Hiroshima 739-8526, Japan}
\altaffiltext{2}{Hiroshima Astrophysical Science Center, Hiroshima University, 1-3-1 Kagamiyama, Higashi-Hiroshima, Hiroshima 739-8526, Japan}
\altaffiltext{3}{U.S.\ Naval Research Laboratory, Code 7653, 4555 Overlook Ave.\ SW, Washington, DC, 20375-5352, USA}
\altaffiltext{4}{Institute of Space and Astronautical Science, JAXA, 3-1-1 Yoshinodai, Chuo-ku, Sagamihara, Kanagawa 252-5210, Japan}
\altaffiltext{5}{Astronomical Observatory, Jagiellonian University, ul. Orla 171, 30-244 Krak\'ow, Poland}

\begin{abstract}

We performed a systematic X-ray study of eight nearby $\gamma$-ray
 bright radio galaxies with {\em Suzaku} {for understanding the
 origin of their X-ray emissions}. The {\em Suzaku} spectra for
 five of those have been presented previously, while the remaining three
 (M\,87, PKS\,0625$-$354, and 3C\,78) are presented here for the first
 time. Based on the Fe-K line strength, X-ray variability, and X-ray
 power-law photon indices, and using additional information on the [O
 III] line emission, we argue for a jet origin of the observed X-ray
 emission in these three sources.  We also analyzed five years of {\em
 Fermi} Large Area Telescope (LAT) GeV gamma-ray data on PKS\,0625$-$354
 and 3C\,78 {to understand these sources within the blazar picture. 
We} found significant 
$\gamma$-ray variability in the
 former object.  Overall, we note that the {\em Suzaku} spectra for both
 PKS\,0625$-$354 and 3C\,78 are rather soft, while the LAT spectra are
 unusually hard when compared with other $\gamma$-ray detected low-power
 (FR\,I) radio galaxies. We demonstrate that the constructed broad-band
 spectral energy distributions of PKS\,0625$-$354 and 3C\,78 are well
 described by a one-zone synchrotron/synchrotron self-Compton model.
 The results of the modeling indicate lower bulk Lorentz factors
 compared to those typically found in other BL Lac objects, but
 consistent with the values inferred from modeling other LAT-detected
 FR\,I radio galaxies. Interestingly, 
the modeling also implies very high peak ($\sim 10^{16}$\,Hz)
 synchrotron frequencies in the two analyzed sources, 
{contrary to previously-suggested scenarios for FR I/BL Lac unification}. 
We discuss the implications of our findings in the context of the FR\,I/BL Lac unification schemes.

\end{abstract}


\section{Introduction}

{Radio galaxies constitute the parent population of blazars, with
low-power radio galaxies thought to be misaligned BL Lacertae objects
(BL Lacs), and higher-power sources thought to be associated with flat
spectrum radio quasars \citep[FSRQs; e.g.,][]{urr95}. In general, the
accretion power determines the radiative properties of the direct
emission of the accreting matter (i.e., the thermal continuum of the
accretion disk and circumnuclear dust), and is reflected in the presence of high-excitation emission lines in the source spectrum \citep{hin79}. In particular, the ``low-excitation radio galaxies'' (LERGs) are considered to be characterized by lower accretion rates (below $\sim 1\%$ in the Eddington units) and radiatively inefficient accretion flows, while ``high-excitation radio galaxies'' (HERGs) are believed to represent high-accretion rate sources with standard (optically-thick, geometrically-thin) accretion disks. The jet power, on the other hand, which in general scales with the total radio luminosity, was proposed to be related uniquely to the large-scale morphologies of radio galaxies \citep{fan74}, with low- and high-power jets forming Fanaroff-Riley (FR) type I and type II structures, respectively. The ``FSRQs/FR\,IIs/HERGs vs. BL Lacs/FR\,Is/LERGs'' unification scenario is not without its caveats, however, as a number of BL Lacs have been found to be associated with FR\,II-like jets and lobes, while some FSRQs display FR\,I large-scale morphologies \citep[e.g.,][]{blu01,hey07,lan08,chi09,kha10}. Also, many FR II-type radio galaxies are classified as LERGs, while some FR Is are known to be hosted by high-excitation nuclei \citep[e.g.,][]{hard06,hard07,hard09,but10,gen13,min14}.} 

{Studying the core emission of radio galaxies in the aforementioned context of unification schemes for active galactic nuclei (AGN) can be challenging however, due to the inevitable contributions from relativistic jets, host galaxies, accretion disks, and disk coronae, any one of which may dominate the observed radiative output of a source in different frequency ranges. Hence, a detailed multifrequency analysis is needed to disentangle robustly various emission components in a number of objects, before drawing any definite conclusions regarding the corresponding jet and accretion luminosities. We note that although} the extended, $\geq$\,kpc-scale jets have been resolved in the X-rays and optical for a number of sources\footnote{See {\tt http://hea-www.harvard.edu/XJET/} and {\tt http://astro.fit.edu/jets/} for continually updated lists of large-scale jets resolved in X-rays and optical, respectively.}, sub-kpc scale structures cannot be imaged at frequencies higher than radio, with a few exceptions \citep[see][]{harr09,goo10,wor10,mey13}. For this reason, even for the brightest radio galaxies such as Cen\,A and NGC\,1275, the origin of the observed optical and X-ray core fluxes is still an open issue \citep[e.g.][]{yam13}. Often, the jet origin of the unresolved core emission is claimed based solely on the modeling of broad-band spectral energy distributions (SEDs; see, e.g., \citealt{chi03} and \citealt{fos05} for the case of NGC\,6251), and therefore the alternative possibilities, such as a disk/corona emission, cannot be ruled out.

In the X-ray domain there are three pieces of evidence that can help to distinguish between thermal disk/corona emission and non-thermal radiation from an unresolved jet: variability, the Fe-K line, and {--- to a lesser extent ---} the spectral slope.  Non-thermal jet emission is expected to be variable on shorter timescales and with greater magnitude than variations in an accretion disk/corona, so fast variability would be an indication of a jet origin, although a lack of such variability would not rule out a jet origin. {Variability constrains were used for example} in the {\em Suzaku} study of Cen\,A by \citet{fuk11b}, who claimed a dominant jet contribution at energies above 100\,keV. \citet{yam13}, on the other hand, concluded that the X-ray core emission from NGC\,1275 is dominated by accreting matter due to a large equivalent width (70--120\,eV) of the detected Fe-K line. Good constraints on Fe-K lines can be achieved by the {\em Suzaku} or {\em XMM-Newton} telescopes, which provide quality X-ray spectra with high signal-to-noise ratios. Obviously, it is possible that in many or even the majority of cases the observed X-ray core emission is a combination of disk/corona and jet contributions, as suggested by the fact that the X-ray spectra of non-jetted AGN are typically harder than those of jetted sources \citep[e.g.,][]{per02}; we note however that the spectral slopes of X-ray continua alone do not provide conclusive evidence in this context {\citep[see the related discussion in][]{hard99}.} Optical line emission can be also used as an additional constraint to diagnose the disk emission in the objects we study.

{During the last decade, rapid developments in $\gamma$-ray astronomy has opened a new window for studying the unification of jetted AGN.} Blazars are generally bright in the $\gamma$-ray band {\citep{hart99,ack11} } due to relativistic beaming of the jet emission, which is expected to not be as significant for misaligned radio galaxies.  Still, recent observations by the {\em Fermi} Large Area Telescope (LAT) in the high-energy (HE; $0.1-100$\,GeV) band, as well as by the atmospheric Cherenkov telescopes in the very high-energy (VHE; $>0.1$\,TeV) band, have revealed that radio galaxies are high-energy emitters. In particular, {\em Fermi}-LAT has detected 11 radio galaxies with 15 months of sky survey data \citep{abd10b}, including M\,87 \citep{abd09b}, Cen\,A \citep{abd10a}, and NGC\,1275 \citep{abd09a}. These three objects are also the only established TeV emitting radio galaxies \citep[][respectively]{aha06,aha09,ale12}, taking into account that the VHE-detected IC\,310 \citep{ale10} is now proposed to be re-classified as a BL Lac object \citep{kad12}.

Here, we report the {\em Suzaku} \citep{mit07} X-ray study of eight nearby (redshifts $z < 0.06$) $\gamma$-ray emitting radio galaxies which are included in the 15-month {\em Fermi}-LAT `misaligned AGN' list \citep{abd10b}.  All these radio galaxies are of the FR\,I type, with the exception of 3C\,111 and NGC\,6251 which display classical FR\,II and intermediate FR\,I/FR\,II large-scale radio morphologies, respectively \citep[e.g.,][]{gli04,sam04,gra12}. The remaining three radio galaxies from the misaligned AGN sample are all distant ($z>0.25$) so we do not explore them here. In Section 2 we present our original {\em Suzaku} data analysis for M\,87, \pks, and \tc; {\em Suzaku} results for the other sources we discuss, 3C\,111, 3C\,120, Cen\,A, NGC\,1275, and NGC\,6251, are quoted from the literature. In Section 3 we also present the analysis of 5 years of {\em Fermi}-LAT data for two particularly intriguing sources, \pks\ and \tc. In Section 4 we discuss the origin of the X-ray emission detected with {\em Suzaku} from unresolved cores of eight analyzed radio galaxies. There we also present for the first time the broad-band SED modeling of \pks\ and \tc\ which appear to have X-rays originating from non-thermal jet emission and have not been previously modeled; the modeling is then compared with the analogous modeling of Cen\ A, NGC\,1275, M\, 87, and NGC\,6251 presented previously in the literature. Our main conclusions are summarized in Section 5.

\section{ {\em Suzaku} Observations }

\subsection{Data Reduction}

{\em Suzaku} is an X-ray observatory which contains two instruments; the X-ray Imaging Spectrometer \citep[XIS;][]{koy07} and the Hard X-ray Detector \citep[HXD;][]{tak07,kok07}. The former consists of four CCD cameras. One CCD has been lost in 2006 November, and thus most of observational results shown in this paper were based on the data of three CCD cameras. The latter consists of PIN photo-diodes and GSO scintillators, surrounded by active shields of BGO scintillators. All the available {\em Suzaku} data for the GeV emitting FR I radio galaxies are summarized in Table\,\ref{obslog}. Results for some of the targets have already been published, as indicated in the table; the {\em Suzaku} results for M\,87, \pks, and \tc\ are presented here for the first time. The Fe-K line equivalent widths and the X-ray luminosities of all the objects provided in Table\,\ref{obslog} and discussed in detail below in this paper are estimated by analyzing the archival {\em Suzaku} data, including the results of \citet{fuk11a}.  All the observations were performed in the XIS $5 \times 5$ or $3 \times 3$ modes, and with the normal mode of the HXD. We utilized data processed with version 2.0--2.7 of the pipeline {\em Suzaku} software, and performed the standard data reduction: a pointing difference of $<1.5^{\circ}$, an elevation angle of $>5^{\circ}$ from the earth rim, a geomagnetic cut-off rigidity (COR) of $>$6 GV, and $>256$ s spent in the South Atlantic Anomaly (SAA). Further selections were applied: Earth elevation angle of $>20^{\circ}$ for the XIS, Cut-off-rigidity (COR) $>$8 GV and the time elapsed from the SAA (T\_SAA\_HXD) was selected to be $> 500$\,s for the HXD. The XIS response matrices were created with {\tt xisrmfgen} and {\tt xissimarfgen} \citep{ish07}. The XIS detector background spectra were extracted at 4--6 arcmin from the target object and then subtracted. We utilized the HXD responses provided by the HXD team. The ``tuned'' ({\tt LCFIT}) HXD background files \citep{fuk09} were used, and the good time interval (GTI) was determined by taking the logical ``AND'' of GTIs among XIS data, HXD data, and HXD background data. For the XIS and HXD-PIN, the Cosmic X-ray Background (CXB) was added to the background spectrum, assuming the flux and spectra in \citet{bol87}, although it was negligible for the HXD-GSO.

\begin{table}[!t]
{\scriptsize
\begin{center}
\caption{Summary of {\em Suzaku} observations of eight LAT-detected FR\,I radio galaxies}
\label{obslog}
\vspace{0.2cm}
\begin{tabular}{cccccc} \hline
\hline
Source & Redshift & ObsID & Date & Exposure$^{\star}$ & References$^{\ddagger}$ \\
\hline
\tc/NGC\,1218 & 0.029 & 706013010 & 2011-08-20 & 97\,ks & this study \\
3C\,84/NGC\,1275 & 0.018 & ---$^{\ast}$ & 2006--2011 & ---$^{\ast}$ & Y13 \\
3C\,111 & 0.049  & 7050400[1-3]0 & 2010-09-02,09,14 & 170\,ks & T11 \\
3C\,120 & 0.033  & 7000010[1-4]0 & 2006-02,03$^{\ast\ast}$ & 147\,ks & K07 \\
\pks\  & 0.055 & 706014010 & 2011-11-03 & 110\,ks & this study \\
M\,87/3C\,274    & 0.004 & 801038010 & 2006-11-29 & 98\,ks & this study \\
Cen\,A/NGC\,5128   & ---$^{\dagger}$ & 100005010, & 2005-08-19, & 94\,ks & M07 \\
& & 7040180[1-3]0 & 2009-07-20/08-05/08-14 & 118\,ks & F11b \\
NGC\,6251        & 0.024 & 705039010 & 2010-12-02 & 87\,ks & E11\\
\hline
\end{tabular}
\end{center}
$^{\star}$ XIS-0 exposure.\\ 
$^{\ast}$ Twelve pointings.\\ 
$^{\ast\ast}$ Four pointings.\\ 
$^{\dagger}$ The assumed distance 3.8\,Mpc.\\ 
$^{\ddagger}$ References: \citet{yam13,tom11,kat07,mar07,fuk11b,eva11}.
}
\end{table}

For XIS and PIN detectors, the energy ranges of 0.45--10\,keV and 17--50\,keV, respectively, were used in the fitting. In addition, we ignored the 1.75--1.88\,keV energy interval in the XIS spectra to avoid the response uncertainty. The X-ray spectra were binned for least $\chi$-square spectral fitting {so that one spectral bin contains more than 20 photons}. XIS photons were accumulated within 4 arcmin of the galaxy center, with the XIS-0 and -3 data co-added for \pks\ and \tc. For M\,87, the XIS-2 CCD was utilized in the same way as above. Since M\,87 is embedded in the bright extended X-ray emission of the Virgo intracluster medium, we took the integration radius as 1 arcmin, and the background spectrum was taken from the 1.5--2.5 arcmin ring, and ignored the HXD-PIN data. Since the GSO signal was not significant in all the analyzed objects, and the resulting upper limits above 40\,keV are not constraining, below we do not discuss the GSO data analysis results. A relative normalization between the XIS-F and XIS-B was left free to vary, while that between {the XIS and PIN was fixed\footnote{\tt http://www.astro.isas.jaxa.jp/suzaku/analysis/hxd/gsoarf2/}  to 1.17.}

\subsection{Results}

\begin{figure}[!t]
\begin{center}
\includegraphics[scale=0.32]{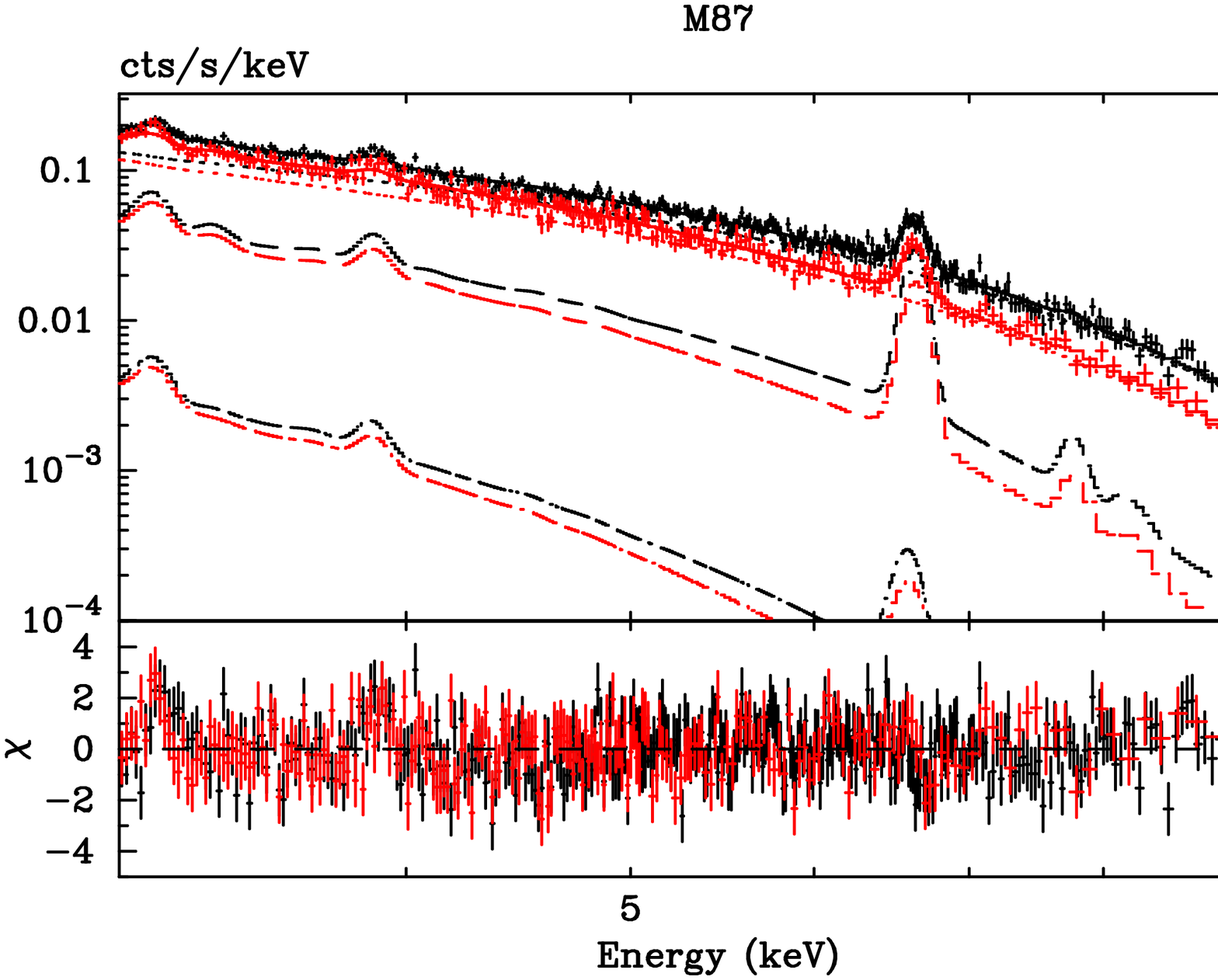}
\includegraphics[scale=0.32]{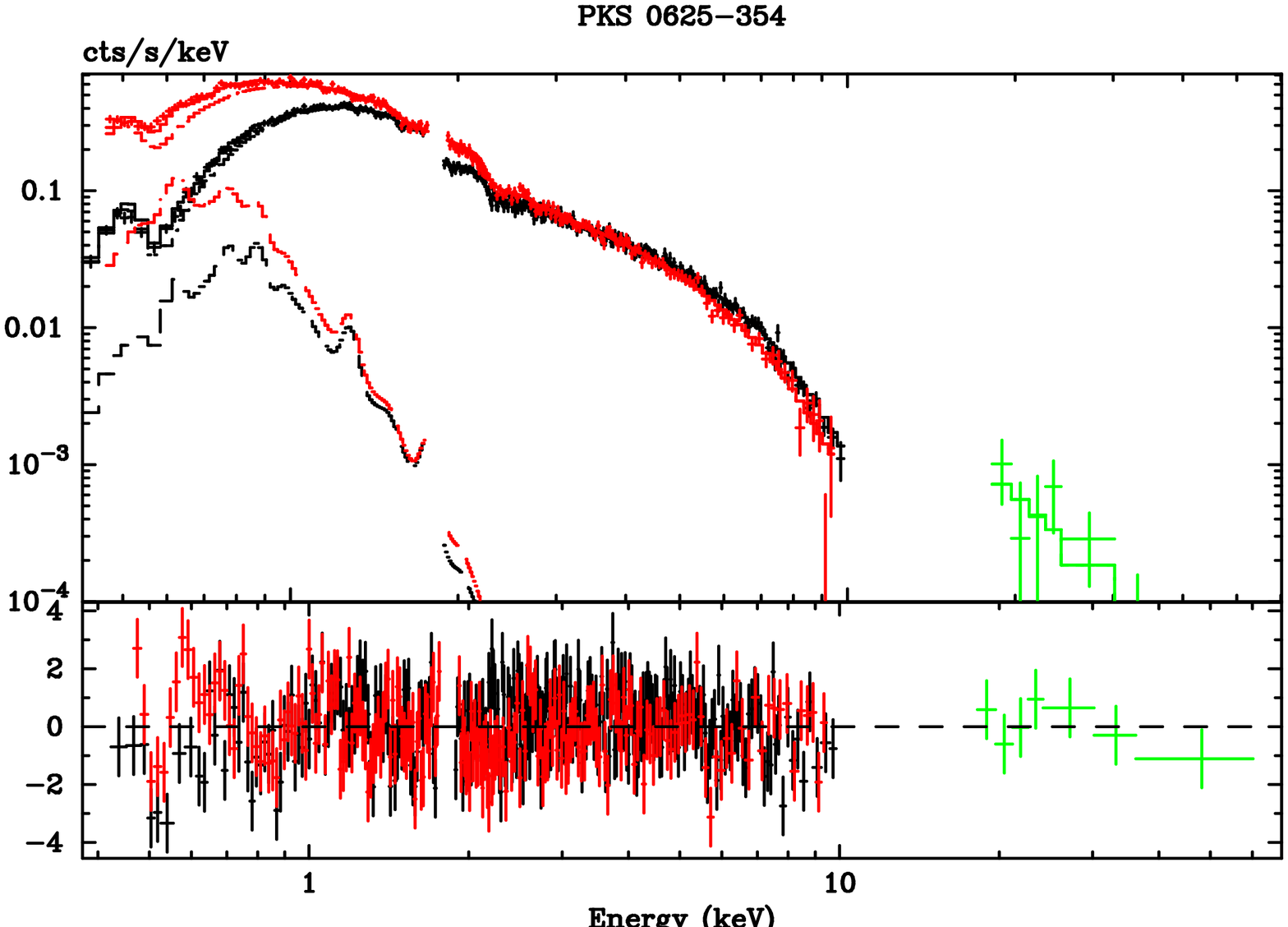}
\includegraphics[scale=0.32]{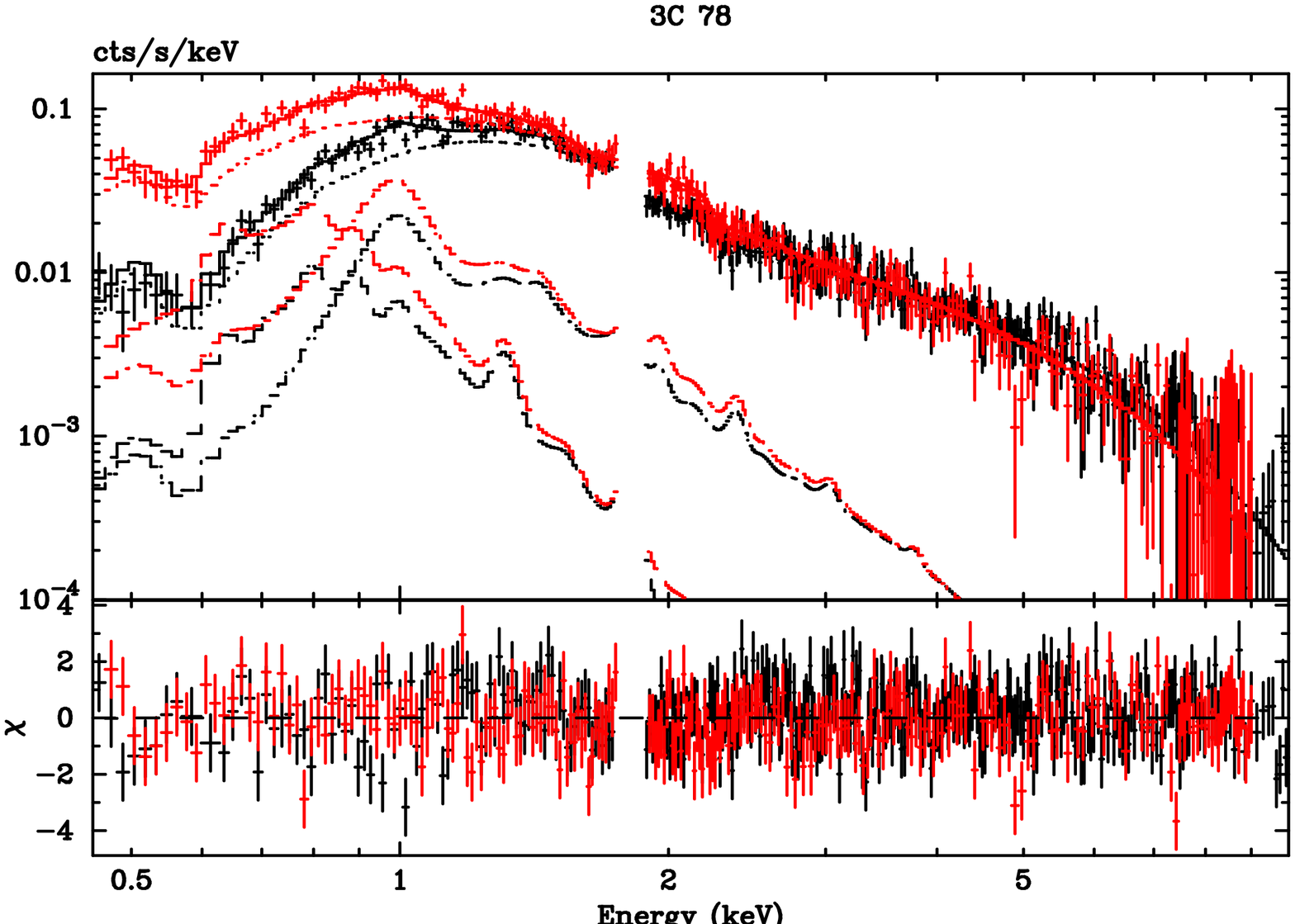}
\vspace{0.5cm}
\caption{{\em Suzaku} spectra of M\,87, \pks, and \tc. The black, red, and green symbols are XIS-F, XIS-B, and HXD-PIN spectra, respectively. The solid line represents the best-fit total model, while the dashed and dotted lines are the {\tt apec} and power-law model components, respectively. We include two {\tt apec} model for M\,87 and \tc. The bottom panels show the residuals in units of $\sigma$.}
\label{xspec}
\end{center}
\end{figure}

M\,87, \pks, and \tc\ are clearly detected with the XIS below 10\,keV; \pks\ is also detected with HXD-PIN above 10\,keV. At first, the obtained {\em Suzaku} X-ray spectra were fit with a single power-law model (PL) multiplied by the Galactic absorption with the column densities fixed to the corresponding values provided by \citet{kal05}. The spectra of all three objects showed some residuals in the soft X-ray band. These residuals could be due to the thermal emission from the hot interstellar or intracluster medium, and/or absorption columns in excess of the Galactic values. Thus, one or two {\tt apec} thermal plasma models were included in the second step of the fitting procedure, and the absorption column densities were let free. In the cases of \pks\ and \tc\ the metal abundance of {\tt apec} was fixed to 0.3 solar, which is typical for galaxy groups \citep[e.g.,][]{fuk04}. In the case of M\,87, good photon statistics allowed us to leave the metal abundance free, but because the temperature structure of the M\,87 core region is complex \citep[e.g.,][]{mat02}, even a two-temperature {\tt apec} model could not fit the soft X-ray continuum well.  Since the detailed investigation of the M\,87 temperature structure is beyond the scope of this paper, we simply applied a two-temperature {\tt apec} model to the 0.7--10\,keV range, constraining the main thermal plasma parameters; then the resulting {\tt apec} model parameter values were fixed during a {\tt apec} + PL model fit to the 3--10\,keV range. Inclusion of the additional thermal components improved the fits, which are summarized in Table\,\ref{xfit} and presented in Figure\,\ref{xspec}.

\begin{table}[!t]
{\scriptsize
\begin{center}
\caption{Summary of {\em Suzaku} data spectral fitting}
\label{xfit}
\vspace{0.2cm}
\begin{tabular}{ccccccccc} \hline
\hline
Source & $N_{\rm H}$ & $kT$ & $Z$ & $L_{\rm 0.5-10\,keV}$ & $\Gamma_{\rm X}$ & $L_{\rm 2-10\,keV}$ & EW &  $\chi^2$/d.o.f \\
 & $10^{20}$\,cm$^{-2}$ & keV & $Z_{\odot}$ & $10^{42}$\,erg\,s$^{-1}$ & & $10^{42}$\,erg\,s$^{-1}$ & eV & \\
 (1) & (2) & (3) & (4) & (5) & (6) & (7) & (8) & (9) \\
\hline
M\,87 & 10$\pm$6 (1.9) & 1.79$\pm$0.02  & 1.4$\pm$0.8 & 13 & 2.42$\pm$0.03 & 0.7 & $<13$ & 674/572 \\
& & 2.28$\pm$0.03 & 1.4$\pm$0.2 \\
\pks\ & 9$\pm$1 (6.36) & 0.24$\pm$0.02 & 0.3$^{(f)}$ & 9.6 & 2.25$\pm$0.02 & 49 & $<7$ & 640/486 \\
\tc\ & 14$\pm$2 (9.51) & 0.29$\pm$0.04 & 0.3$^{(f)}$ & 1.0 & 2.32$\pm$0.04 & 2.0 & $<75$ & 572/567 \\
& & 1.07$\pm$0.06 & 0.3$^{(f)}$ \\
NGC\,6251 & & & & & 1.82$_{-0.05}^{+0.04}$ & 2.86 & $<66$ & \\
3C\,111 & & & & & 1.65$\pm$0.02 & 259 & 40$\pm9$ & \\
3C\,120 & & & & & 1.75$_{-0.02}^{+0.03}$ & 100 & 50$\pm10$ & \\
NGC\,1275 & & & & & 1.73$\pm0.03$ & 7.7 & 75$\pm7$ & \\
Cen\,A & & & & & 1.73$\pm0.03$ & 10 & 76$\pm3$ & \\
\hline
\end{tabular}
\end{center}
(1) Source. (2) Absorption column density for {\tt phabs} model; the values in the parentheses are the Galactic values $N_{\rm H,\,Gal}$ from \citet{kal05}. (3) Temperature in the ${\tt apec}$ fits. For M\,87 and \tc, two temperatures are shown in the two {\tt apec} model. (4) Abundance in the {\tt apec} fit (fixed in the cases of \pks\ and \tc). (5) Absorption-corrected ${\tt apec}$ luminosity. (6) Photon index in the PL fit; {values for bottom five objects are quoted from the corresponding references in Table\,\ref{obslog}}. (7) Absorption-corrected PL luminosity. (8) Equivalent width of Fe-K line. (9) Goodness of the fit (in the case of M\,87 the provided $\chi^2$/d.o.f value corresponds to the PL fit to the 3--10\,keV range; see section 2.2 for details). Values for bottom five objects in column 7, 8, and 9 are originally derived by us.}
\end{table}

The plasma temperatures and luminosities derived for \pks\ and \tc\ are consistent with those of interstellar and intracluster media found in elliptical galaxies and galaxy groups \citep[e.g.,][]{bir04,fuk06,die07}. We also checked the archival {\em Chandra} X-ray data for \pks\ and confirmed that the X-ray appearance of the source is almost point-like with a faint extended halo due to the host elliptical/poor galaxy cluster Abell\,3392. The temperatures of two {\tt apec} components derived for M\,87 are consistent with those obtained before by \citet{mat02} based on the analysis of {\em XMM-Newton} data; the two distinct thermal components in this case are due to the projection of cool-core and hot-periphery cluster emission at the cluster center.

The absorption column densities $N_{\rm H}$ for the three analyzed radio galaxies are slightly larger than the corresponding Galactic value of \citet{kal05}; see Table\,\ref{xfit}. This might be due to the spectral curvature in the soft X-ray band, but we cannot rule out the uncertainty of the $N_{\rm H,\,Gal}$ database, {spectral modeling dependency of the thermal emission}, and also the intrinsic absorption {by the interstellar medium in host galaxies}. The derived power-law photon indices $\Gamma_{\rm X}$, distributed within a narrow range of 2.22--2.45, are somewhat steeper but not extraordinary for coronal emission of Seyfert galaxies. They are also consistent with the X-ray photon indices of high-peaked BL Lac objects \citep[e.g.,][]{don05,aje09}, i.e. the aligned counterpart to FR\,I radio galaxies, where the X-rays have a jet origin. Fluorescence Fe-K lines are common features in AGNs dominated by disk emission. However, none of the objects analyzed here show significant fluorescence Fe-K lines, except for the ionized Fe-K lines from the hot plasma around the M\,87 core.  We obtained upper limits of the equivalent widths (EWs) of the narrow Fe-K lines at the rest-frame energy of 6.4\,keV (see Table\,\ref{xfit}); these are particularly low in the cases of \pks\ and M\,87.  

\begin{figure}[!t]
\begin{center}
\includegraphics[scale=0.32]{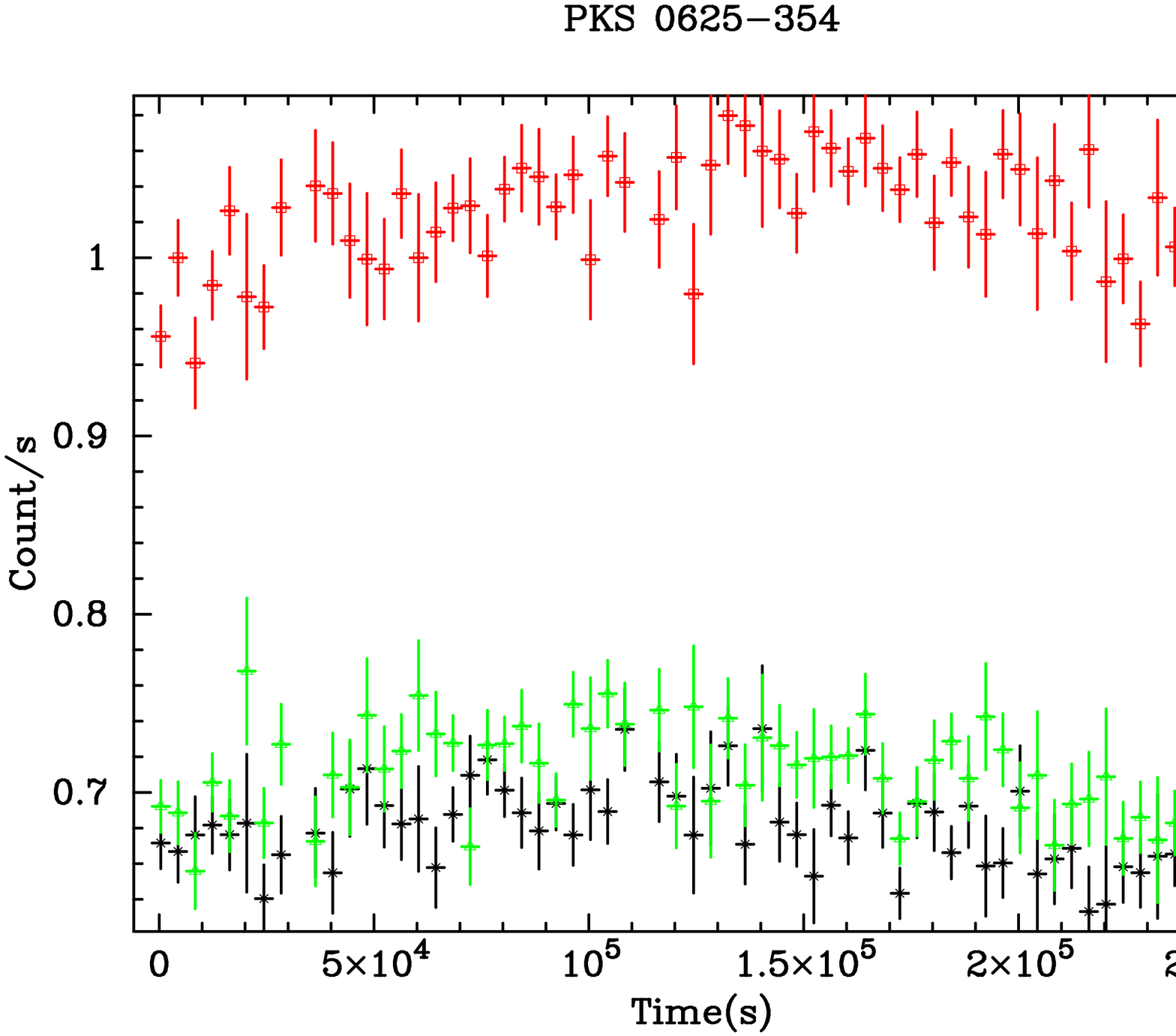}
\includegraphics[scale=0.32]{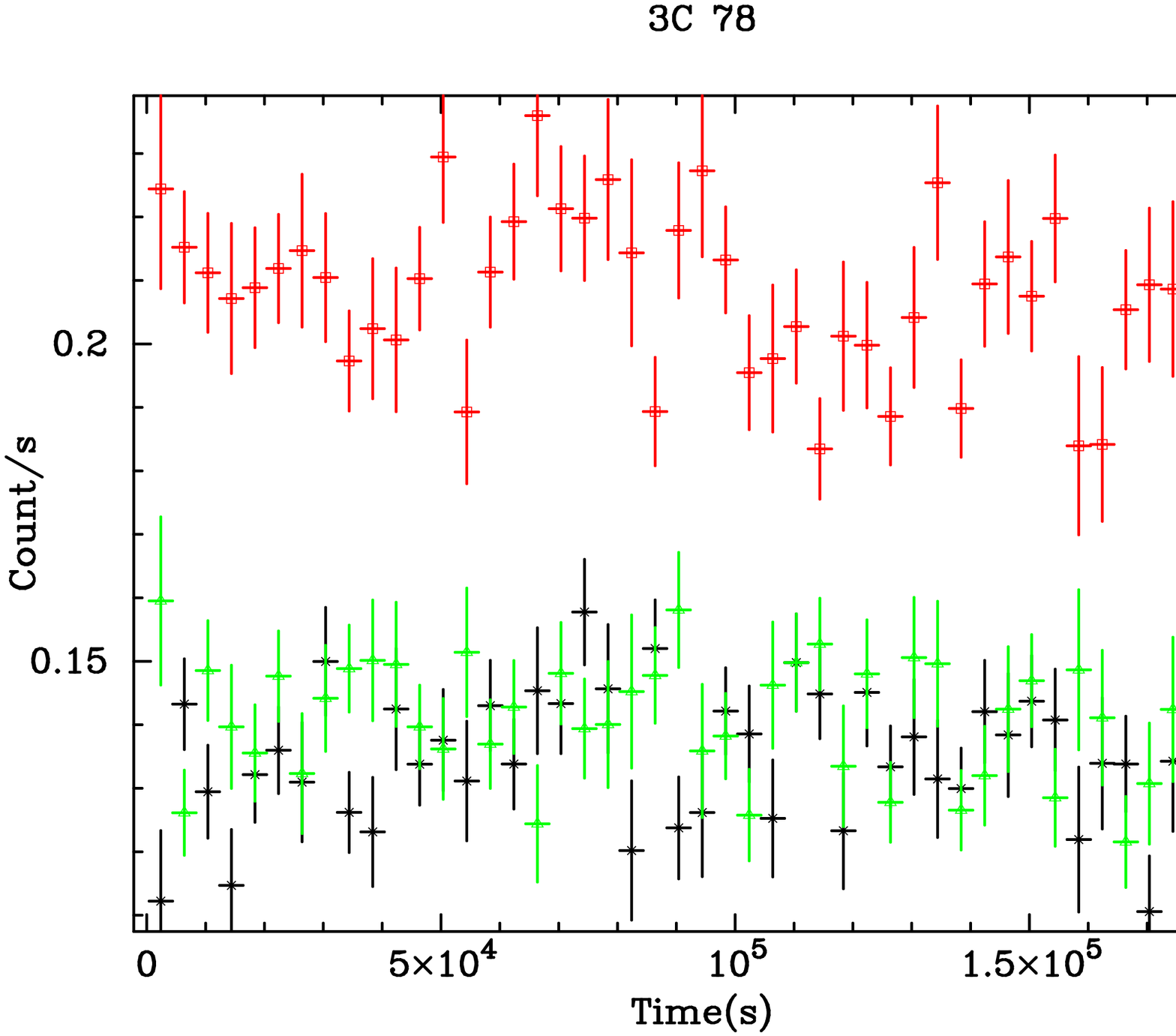}
\vspace{0.5cm}
\caption{{\em Suzaku} X-ray light curves of \pks\ (left) and \tc\ (right) in the 0.45--8\,keV band. The size of the time bins is 4000\,s. The red data points are XIS-B, and others are XIS-F. }
\label{xlc}
\end{center}
\end{figure}

\citet{gli08} reported on the 2005 {\em XMM-Newton} data analysis for \pks. They found a power-law X-ray component with a photon index of 2.52$_{-0.03}^{+0.02}$ and a flux of $2.6 \times 10^{-12}$ erg\,cm$^{-2}$\,s$^{-1}$; the EW of the Fe-K line was constrained to be $<182$\,eV. \citet{tru99} reported on {\em BeppoSAX} observations of \pks\ and \tc\ from 1996--1997; assuming a photon index of 2.3 for both sources, they derived the 1--10\,keV luminosities of the power-law components as $1.8 \times 10^{43}$\,erg\,s$^{-1}$ and $1.5 \times 10^{42}$\,erg\,s$^{-1}$,
respectively. Our {\em Suzaku} observations reveals therefore a flatter-spectrum and brighter (by a factor of 2--3) X-ray emission of \pks\ when compared with the previous epochs. For \tc, the X-ray flux is almost the same as that in 1997, and we constrain the power-law photon index and Fe-line EW for the first time. The Fe-K line EWs are in general much more strongly constrained in this study than ever before.

We also investigated the X-ray time variability of the analyzed radio galaxies during the {\em Suzaku} observations; see Figure\,\ref{xlc}. No statistically significant variability was found in the 0.45--8\,keV range for \tc, however most likely only due to a very low photon statistics. \pks, on the other hand, showed a small amount of variability with a time scale of 1--2 days and an amplitude of $\sim 10\%$. {Significant X-ray variability of M\,87 detected in the acquired {\em Suzaku} data will be discussed \citep[and compared with ongoing {\em Chandra} monitoring; see][]{harr09} in the forthcoming paper.}

\section{ {\em Fermi}-LAT Observations}

The {\em Fermi}-LAT is a pair conversion telescope which has a field of view of about 20\% of the sky from 20\,MeV to over 300\,GeV \citep{atw09}. Since our results indicate that the X-ray spectra of \pks\ and \tc\ are dominated by jet emission, we analyzed five years of LAT data for those two radio galaxies. As mentioned in Section\,1, the SED modeling of M\,87 was preformed by \citet{abd09b}.

\subsection{Data Analysis and Localization}

We analyzed the LAT {\tt P7REP} data from 2008 August 4 to 2013 August 4, corresponding to mission elapsed time (MET) 239557420 to 397353600. Source class ({\tt evclass=2}) events were selected with a zenith angle cut of $<$100$^{\circ}$, and a rocking angle cut of 52$^{\circ}$.  For the analysis, LAT Science Tools version v9r32p5 was utilized with the {\tt P7REP\_SOURCE\_V15} Instrument Response Functions (IRFs). Both radio galaxies are clearly visible in the 0.2 to 300\,GeV LAT counts maps. We obtained a localization of the $\gamma$-ray sources associated with each galaxy with the {\tt gtfindsrc} task.  The resulting localizations were reduced to the 95\% confidence localization error $r_{95}$= 0.042$^{\circ}$ centered at (RA, DEC) = (96.785$^{\circ}$, $-35$.488$^{\circ}$) for \pks\ (NED: 96.778$^{\circ}$, $-35$.488$^{\circ}$), and $r_{95}$= 0.089$^{\circ}$ centered at (47.145$^{\circ}$, 4.130$^{\circ}$) for \tc\ (NED: 47.109$^{\circ}$, 4.111$^{\circ}$). These localized positions are consistent within 0.007$^{\circ}$ and 0.046$^{\circ}$ from the center {of the two targets, respectively}.

\subsection{Results}

\begin{table}[!b]
{\scriptsize
\begin{center}
\caption{Summary of the {\em Fermi}-LAT data spectral fitting}
\label{gfit}
\vspace{0.2cm}
\begin{tabular}{cccccc} \hline
\hline
Source & $\Gamma_{\rm HE}$ & $F_{\rm 0.1-100\,GeV}$ & TS & GBMN & IBMN \\
(1) & (2) & (3) & (4) & (5) & (6) \\
\hline
\pks\ & 1.72$\pm$0.06 & 6.7$\times10^{-9}$ & 403.2 & 1.06$\pm$0.02 & 1.38$\pm$0.03 \\
\tc\ & 2.01$\pm$0.16 & 4.9$\times10^{-9}$ & 61.3 & 1.04$\pm$0.01 & 0.95$\pm$0.03 \\
\hline
\end{tabular}
\end{center}
(1) Source. (2) HE $\gamma$-ray photon index. (3) Photon flux in the units of ph\,cm$^{-2}$\,s$^{-1}$. (4) Test Statistics of the detection. (5) Galactic background model normalization. (6) Isotropic background model normalization.
}
\end{table}

We extracted the data within a 12$\times$12\,deg$^2$ rectangular region centered on each object.  The binned likelihood fitting with the {\tt gtlike} tool was performed.  The field background point sources within 14.5$^{\circ}$ from each source, listed in the LAT 2 year catalog \citep{nol12}, were included, and their spectra were assumed to be power-laws with the photon indices fixed to the catalog values. The standard LAT Galactic emission model was used ({\tt gll\_iem\_v05.fits}) and the isotropic diffuse gamma-ray background and the instrumental residual background were represented as a uniform background ({\tt iso\_source\_v05.txt})\footnote{These background models are available at the FSSC:\\ {\tt http://fermi.gsfc.nasa.gov/ssc/data/access/lat/BackgroundModels.html}}. A likelihood analysis was performed with the energy information binned logarithmically in 30 bins in the 0.2--300\,GeV band, and the spatial information binned with 0.15$\times$0.15\,deg$^2$ bin size. For the Galactic and isotropic emission models the normalizations were left free. The spectra of both the analyzed radio galaxies were modeled as power-laws.

\begin{figure}[!t]
\begin{center}
\includegraphics[scale=0.32]{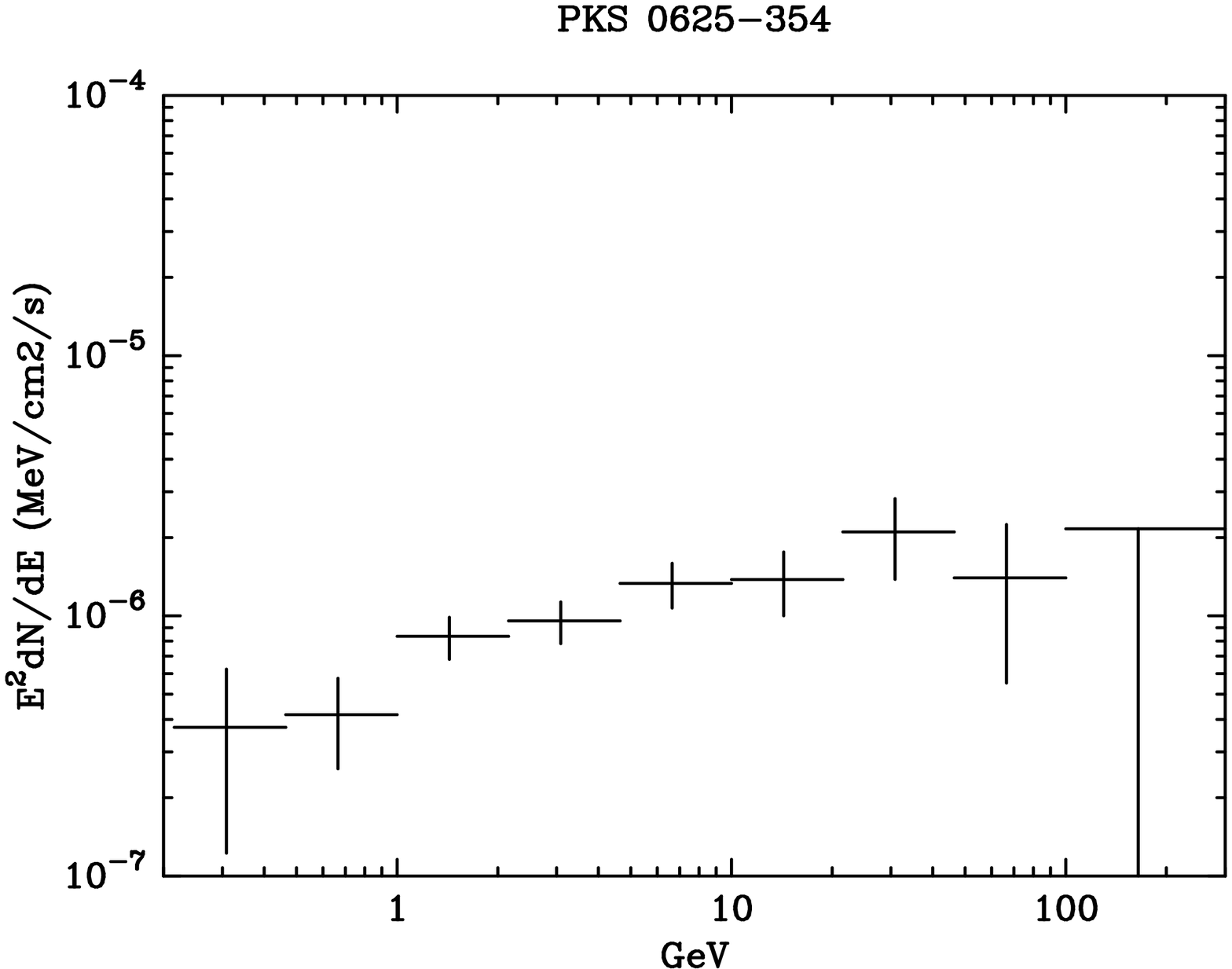}
\includegraphics[scale=0.32]{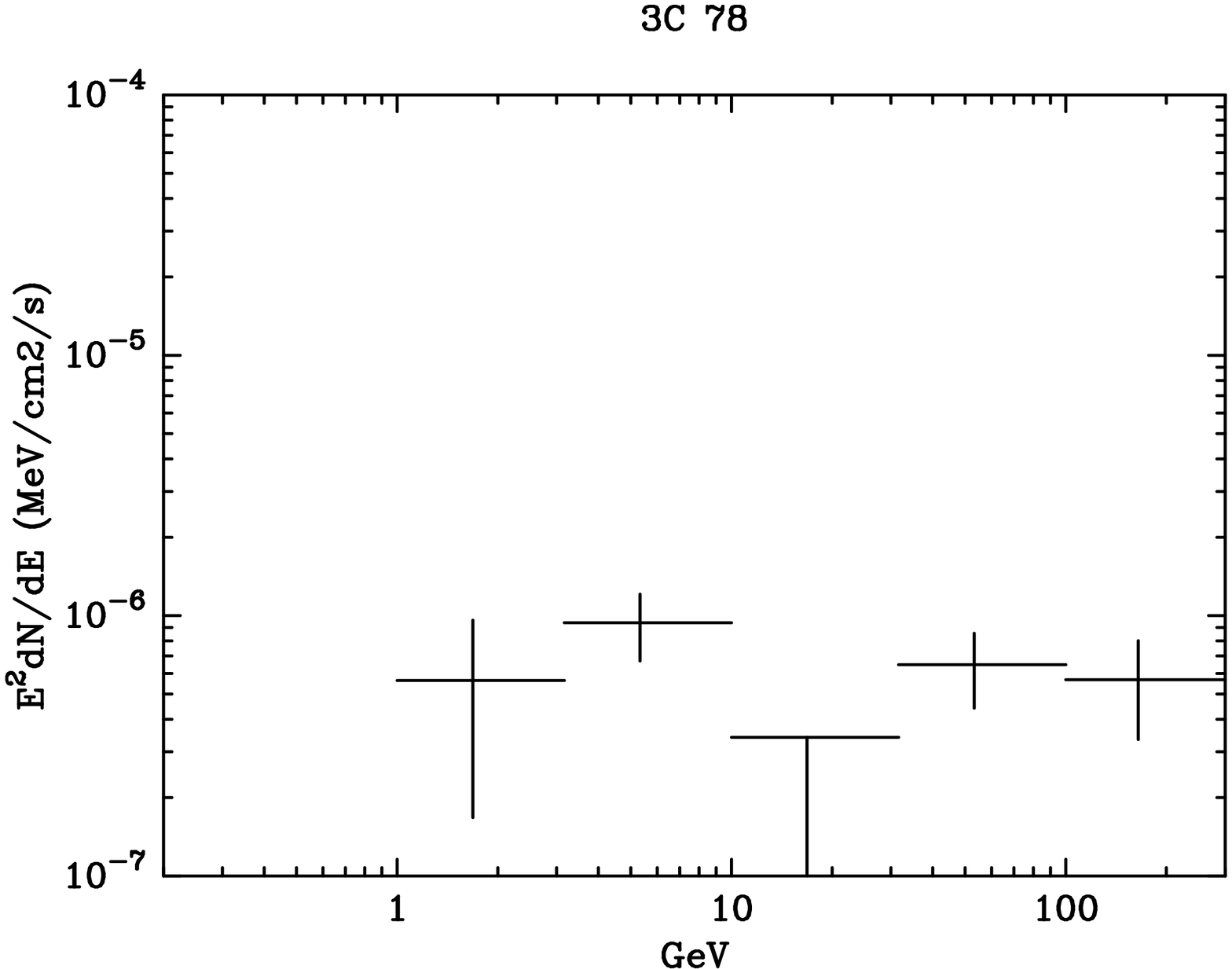}
\vspace{0.25cm}
\caption{{\em Fermi}-LAT $\gamma$-ray spectra of \pks\ (left) and \tc\ (right). The data points with only the lower error bar represent upper limits of 90\% confidence level flux.}
\label{gspec}
\end{center}
\end{figure}

Table\,\ref{gfit} summarizes the {\em Fermi}-LAT data fitting results. The HE $\gamma$-ray photon indices $\Gamma_{\rm HE}$ (evaluated for the 0.2--300\,GeV interval) are within the `standard' blazar range (1.3--3.0) \citep{abd10d}, and we note that \pks\ and \tc\ have the hardest LAT spectra among the entire 15 month `misaligned AGN' sample \citep{abd10b}. The values of $\Gamma_{\rm HE}$ and photon fluxes $F_{\rm 0.1-100\,GeV}$ provided here are in agreement with those given in the 15 month catalog.

In order to obtain model-independent spectra in the 0.2--300\,GeV range for our two sources, we performed the {\tt gtlike} spectral analysis for several independent energy bins, which were spaced logarithmically. Nine energy bins were analyzed for \pks\, and six energy bins for \tc. In each energy bin, we fixed the power-law photon index to 2.0.  Figure\,\ref{gspec} shows the resulting spectra, where we do not detect signals below 1 GeV ($TS<5$) for \tc.  Interestingly, the $\gamma$-ray detection is significant up to 100\,GeV for both objects. In addition, the analysis indicates a break in the $\gamma$-ray spectrum of \pks. We therefore applied the broken power-law model with {\tt gtlike}, and found that the likelihood $L$ increased by $2 \Delta \log L = 401.4$, corresponding to $20\sigma$\footnote{TS=$2 \Delta \log L$ is distributed as $\chi^2$ for one degree of freedom.}; the resulting photon indices were then derived as $1.69 \pm 0.07$ and $4.97 \pm 1.53$ below and above the break energy $64 \pm 23$\,GeV, respectively.

\begin{figure}[!t]
\begin{center}
\includegraphics[scale=0.32]{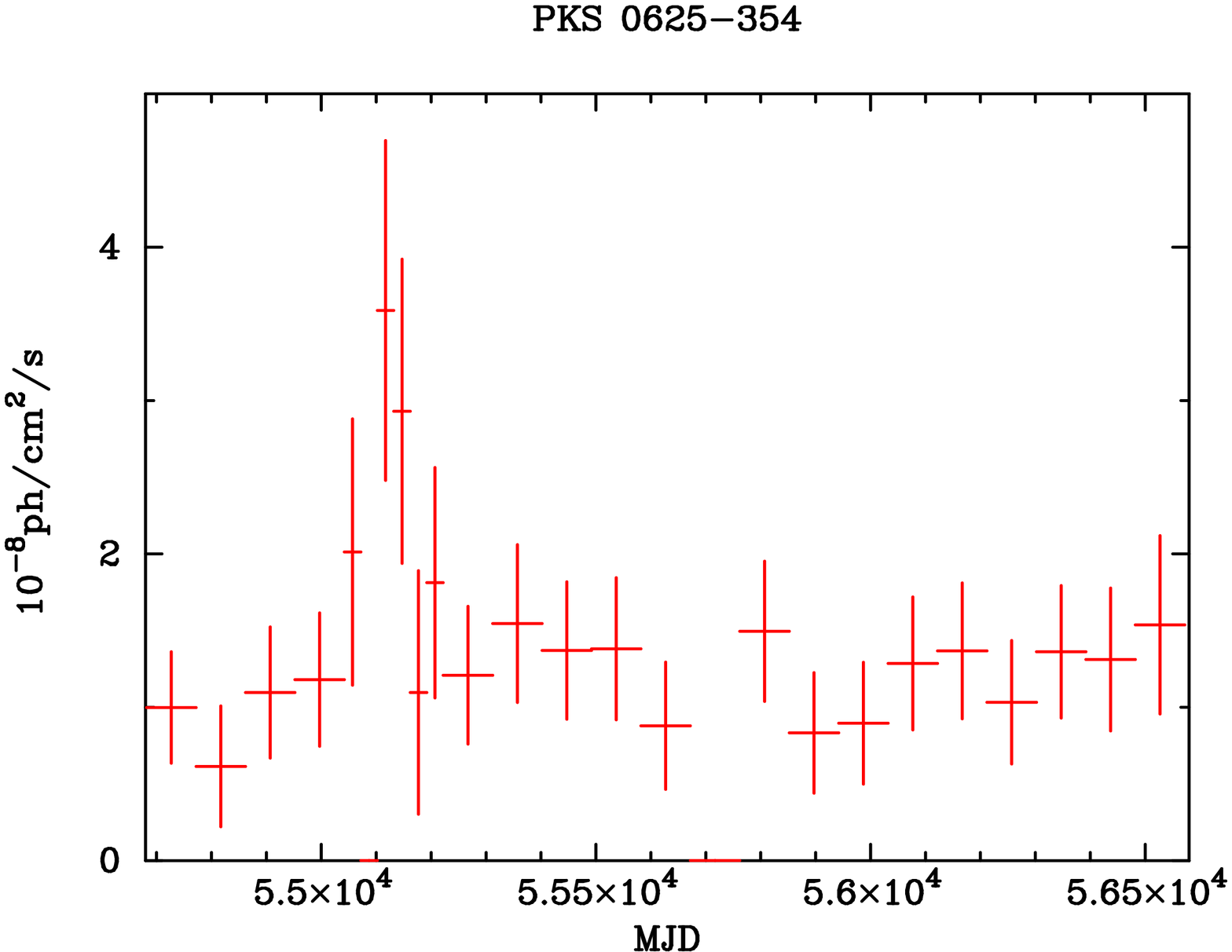}
\includegraphics[scale=0.32]{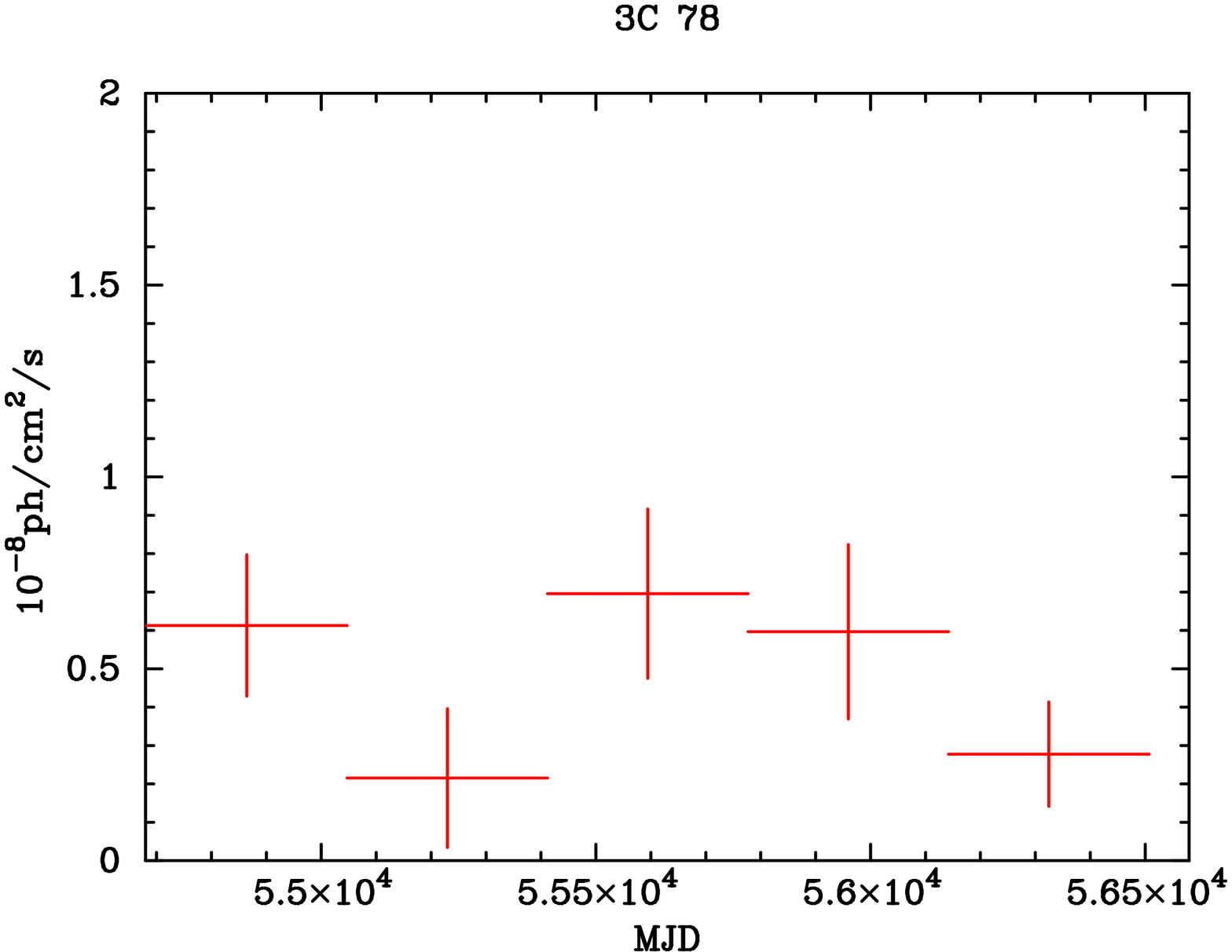}
\vspace{0.25cm}
\caption{{\em Fermi}-LAT 5-year light curves of \pks\ (left) and \tc\ (right) in the 0.2--300\,GeV range.}
\label{glc}
\end{center}
\end{figure}

In order to investigate the $\gamma$-ray variability of the two analyzed radio galaxies, we binned the LAT data into 30- or 90-day bins for \pks, and 1-year bins for \tc. The {\tt gtlike} analysis was performed for each time bin in the same way as the 5-year analysis of the 0.2--300\,GeV band. Figure\,\ref{glc} shows the resulting HE $\gamma$-ray light curves. No significant variability can be claimed for \tc, but \pks\ displayed a rather pronounced flare during the second year of the LAT operation. Keeping in mind the hardening in the HE $\gamma$-ray spectrum of NGC\,1275 detected during the flaring state by \citet{kat10}, we split the LAT data for \pks\ into the two 2.5-year long intervals, and performed the {\tt gtlike} analysis for each epoch separately, but we did not find any significant spectral evolution.

\section{Discussion}

\subsection{Origin of the X-ray Emission}

In this subsection, we summarize {\em Suzaku} X-ray studies of the GeV-emitting FR\,I radio galaxies. Together with the new results on M\,87, \pks, and \tc\ reported in this paper, we refer to {\em Suzaku} results from publications listed in Table\,\ref{obslog}.

For most of the objects in our sample the X-ray spectra are quite
similar to those of {radio-quiet (non-jetted)} Seyfert galaxies, and
only in a few cases the power-law photon indices seem somewhat steeper
than those typically derived for Seyferts \citep[$\Gamma_{\rm X} \sim
1.5$--$2.1$; e.g.,][]{per02}. Therefore, the key feature in
distinguishing between the disk/corona versus jet origin for the
observed X-ray emission is a fluorescence neutral narrow Fe-K line. This
line, commonly observed in Seyfert galaxies {accreting at moderate
and higher rates with $>10^{-2} L_{\rm edd}$, where $L_{\rm edd}$ is
the Eddington luminosity}, is believed to originate
from the Compton-thick {dusty} torus, which subtends the accretion
disk with a large solid angle, as a result of reflection of coronal
nuclear X-ray emission. The Fe-K line width and slow variability support
the torus origin of a narrow Fe-K line \citep{fuk11a}. In the case where
the X-ray emission is dominated by non-thermal jet radiation, one should
not expect a strong Fe-K line, since the jet emission is beamed away
from the disk, and so jet photons are not likely to be reflected by the
torus. {At the same time, sources accreting at particularly low
rates {with $<10^{-2} L_{\rm edd}$} may in principle lack large amounts of circumnuclear dust or prominent coronal components, and as such may be rather weak Fe-K line emitters. Among our sample of targets, four HERGs including Cen\,A, NGC\,1275, 3C\,120, and 3C\,111 reveal clear Fe-K lines, while the objects classified as LERGs --- namely M\,87, \pks, \tc, and NGC\,6251 --- do not.} 

\begin{figure}[!t]
\begin{center}
\vspace{-1cm}
\includegraphics[scale=0.4]{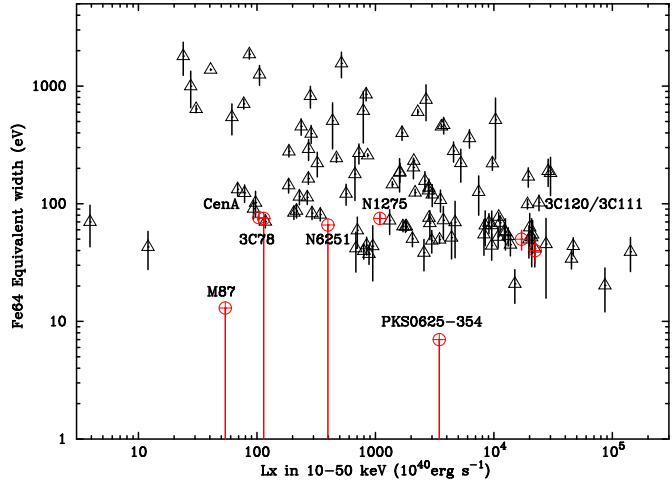}
\vspace{0.5cm}
\caption{Fe-K line EW plotted against the X-ray luminosity for our sample of radio galaxies (red circles) and Seyfert galaxies (black triangles) analyzed by \citet{fuk11a}. The data points with only the lower error bar represent upper limits.}
\label{fek}
\end{center}
\end{figure}

Figure\,\ref{fek} shows the Fe-K EW plotted versus the X-ray luminosity {$L_{\rm X}$ (spanning a wide range from $\lesssim 10^{41}$\,erg\,s$^{-1}$ up to $\sim 10^{46}$\,erg\,s$^{-1}$) obtained with {\em Suzaku} for Seyfert galaxies, together with our sample of radio galaxies. The Fe-K line-detected radio galaxies are found in the same region of the EW --- $L_{\rm X}$ plane as Seyfert galaxies.} Thus, their X-ray emission is likely dominated by disk emission. Other X-ray properties, such as time variability and X-ray relative flux in the SED, also support the disk origin for these sources \citep{fuk11b,yam13,kat11}. However, Cen\,A is characterized by an excess in hard X-rays (above 100\,keV) that smoothly connects to the GeV continuum of the source \citep{fuk11b}. This may imply that for this source the X-ray radiation consists of both thermal disk/corona at low energies and non-thermal jet emission at higher energies.

\begin{figure}[!t]
\begin{center}
\vspace{-1cm}
\includegraphics[scale=0.35]{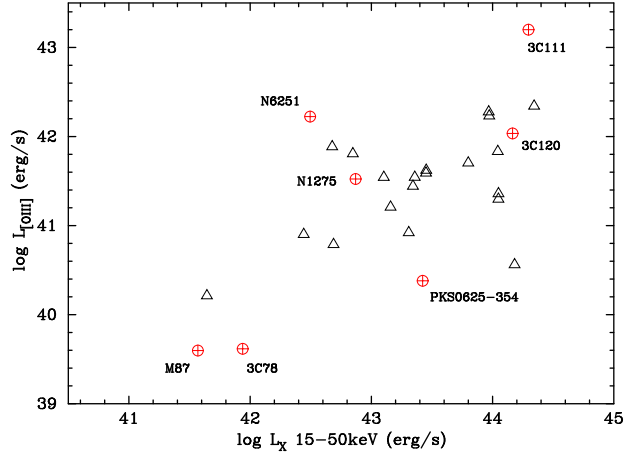}
\vspace{0.5cm}
\caption{Relation between the X-ray luminosity and [O III] 5007\AA\ luminosity for our sample of radio galaxies (filled circles; NED data base) and Seyfert galaxies \citep[empty triangles; data from][]{mul94,win10}.}
\label{o3}
\end{center}
\end{figure}

{\em Suzaku} puts strong upper limits on the Fe-K line EWs for M\,87,
\pks, and \tc. All of these limits are significantly lower than the Fe-K
EWs measured for Seyfert galaxies {with comparable X-ray
luminosities}, suggesting that the X-ray emission of these three sources
is most likely of jet origin. This, in fact, constitutes {the first
strong \emph{indication}} for compact non-thermal jet emission in the
X-ray band for \pks\ and \tc. Note that there is already some evidence
that the X-ray emission from the M\,87 core is dominated by the jet,
albeit different jet regions seem to be pronounced at different activity
level of the source \citep{harr09,harr11}. {One should emphasize
however that non-detection of the Fe-K line in a source spectrum does not prove robustly the dominance (or even a presence) of a jet component.}

Optical [O III] line emission is also a meaningful indicator of pronounced disk emission, since this line is emitted by the extended gas photoionized by strong disk UV emission. Figure\,\ref{o3} shows a plot of the X-ray luminosity versus [OIII] 
luminosity. {The plot reveals some (weak) hints for a $L_{\rm X} -
L_{\rm [O\ III]}$ correlation in the case of Seyferts, but it is not obvious for the analyzed radio galaxies. In particular, the three HERGs in the sample (NGC\,1275, 3C\,111, and 3C\,120) seem to be located in the same region as Seyferts, thus obeying the correlation, while the LERGs seem to be located significantly off the track. This is in agreement with the results of \citet{hard09} and \citet{min14}, who showed that the $L_{\rm X} - L_{\rm [O\ III]}$ correlation persists for HERGs, and is not followed by the LERGs in general. This finding can be considered as further support for the scenario} for the disk/corona emission dominating the X-ray spectra in ``Seyfert-like'' sources NGC\,1275, 3C\,111, and 3C\,120, and the jet emission dominating the X-ray output of the outliers like \pks. However, a caveat for this conclusion is that luminosity-luminosity correlations in flux-limited samples may not be real, but only induced by selection effects.

We have summarized the evidence for the disk/corona versus jet origin for our sample, as discussed in this Section, in Table \ref{diskjetsummary}.

\begin{table}[!t]
{\scriptsize
\begin{center}
\caption{Summary of evidence for disk/corona versus jet origin for X-ray emission}
\label{diskjetsummary}
\vspace{0.2cm}
\begin{tabular}{ccccccc} \hline
\hline
Source & Fe-K line & X-ray spectral index & X-ray variability & [O III] line & Type [ref.]\\
\hline
\tc\ & jet & jet & inconclusive & jet & LERG [B10]\\
3C\,84 & disk/corona & inconclusive & inconclusive & disk/corona & HERG/LERG$^{\dagger}$\\
3C\,111          & disk/corona & inconclusive & inconclusive & disk/corona & HERG$^{\ddagger}$ [E00]\\
3C\,120          & disk/corona & inconclusive & inconclusive & disk/corona & HERG$^{\ddagger}$ [E00]\\
\pks\  & jet & jet & inconclusive & jet & LERG [M14]\\
M\,87    & jet & jet & jet & jet & LERG [G13]\\
Cen\,A           & disk/corona & inconclusive & jet & inconclusive & HERG [E04]\\
NGC\,6251        & jet & inconclusive & inconclusive & jet & LERG [E11]\\
\hline
\end{tabular}
\end{center}
Refs: [B10, E00, M14, G13, E04, E11]\citet{but10,era00,min14,gen13,eva04,eva11}.\\
$^{\dagger}$  3C\,84 is diversely classified in the literature; see, e.g., \citet{hard09,but10,gen13}.
$^{\ddagger}$ 3C\,111 and 3C\,120 are archetype examples of Broad-Line Radio Galaxies (BLRGs).
}
\end{table}

\subsection{X-ray/$\gamma$-ray Connection}

\begin{figure}[!t]
\begin{center}
\vspace{0.5cm}
\includegraphics[scale=0.32]{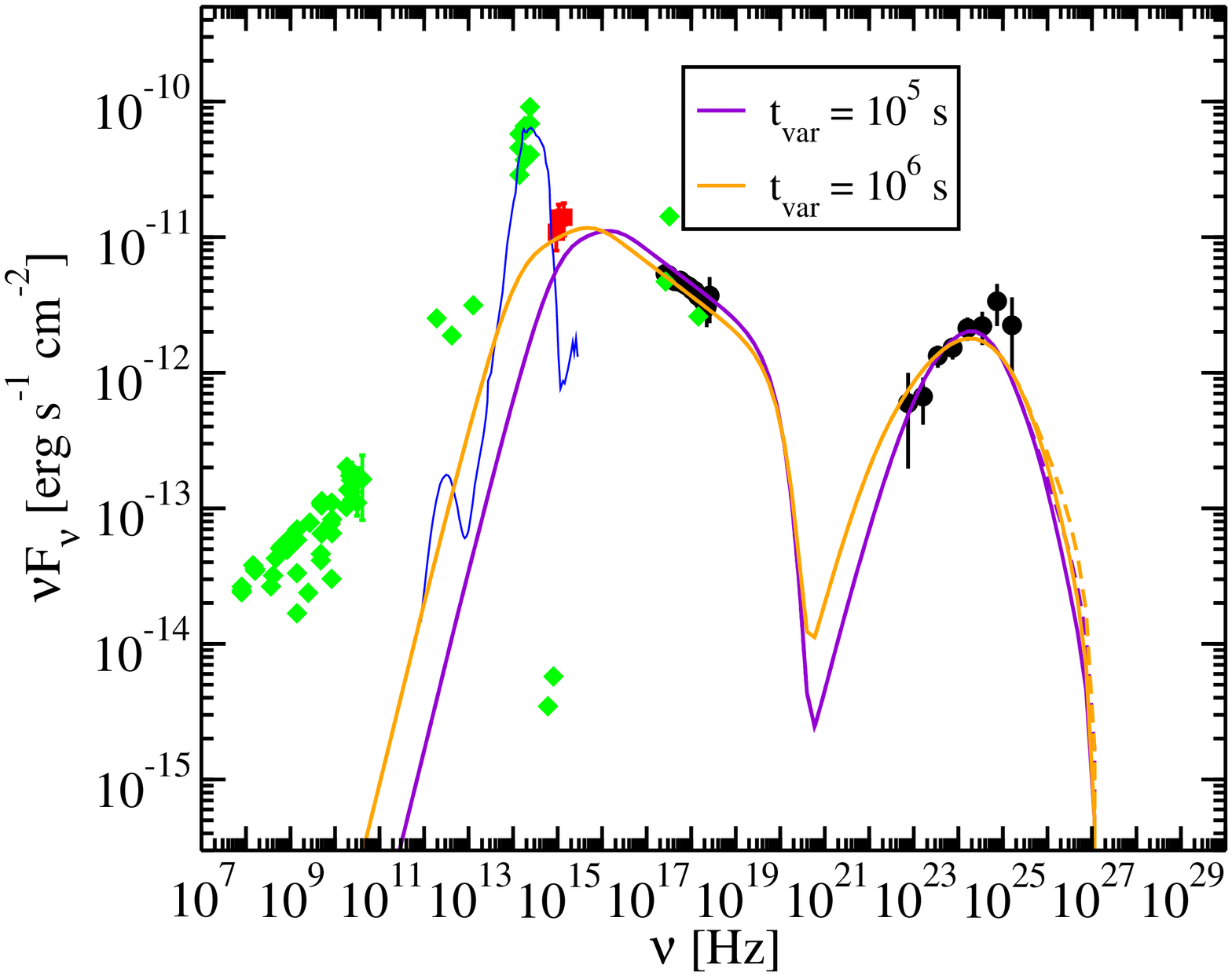}
\hspace{1cm}
\includegraphics[scale=0.32]{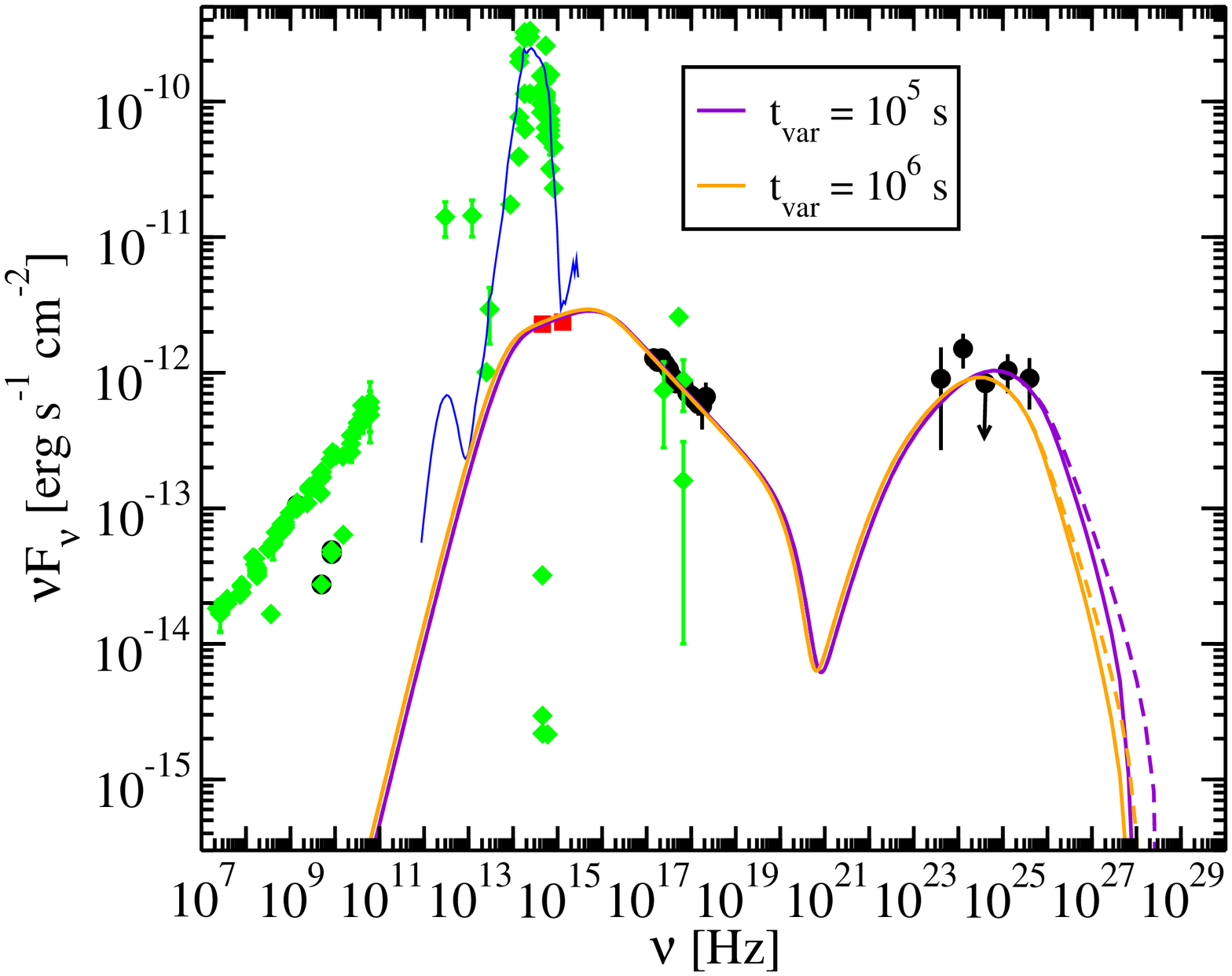}
\caption{SEDs of \pks\ ({left}) and \tc\ ({right}). Black circles indicate the {\em Suzaku} X-ray and {\em Fermi}-LAT $\gamma$-ray data presented in this paper, green diamonds are archival data from NED, and red squares are the {\em XMM-Newton} OM data for \pks\ \citep{gli08} and core {\em HST} data for \tc\ \citep{chi02}. The thick curves denote the synchrotron/SSC model fits with two different variability timescales, as given in the legend. The solid curves include $\gamma\gamma$ absorption with the EBL model of \citet{fin10}, while the dashed curves do not. The thin blue curves are the elliptical galaxy template from \citet{sil98}, adjusted to the redshifts of the sources.}
\label{sed}
\end{center}
\end{figure}

The GeV $\gamma$-ray emission from radio galaxies could originate from the pc/sub-pc scale jet, where the likely mechanism is synchrotron self-Compton (SSC) or external Compton (EC) scattering of the dust torus or broad-line region emission.  It could also originate from EC scattering of CMB photons by electrons in the {100\,kpc-scale jets} or the radio lobes, as established for Cen\,A (Abdo et al. 2010b).  {Somewhat tentative detections of the lobes'  $\gamma$-ray emission have been reported also for the intermediate FR\,I/FR\,II sources NGC\,6251 \citep{tak12} and Cen\,B \citep{kat13}, based on the spatial offset between the best-fit position of the $\gamma$-ray source and the position of the radio core, or the extension of the $\gamma$-ray source aligned to the large-scale radio structure, respectively.}  The $\gamma$-ray emission could not be localized {precisely or potentially resolved for \tc\ or \pks, due to the large position errors and a combination of relatively large source distances (when compared with those of Cen\,A or Cen\,B) and relatively small sizes of their FR\,I-type radio structures. The variability of \pks\ however makes the pc scale origin much more likely for this source, and we favor this interpretation for \tc\ as well.} This also allows us to make a connection between the $\gamma$-rays and X-ray emission from \pks\ and \tc, which was established as likely of jet origin in the previous section. {We therefore model} the broadband SEDs of these two sources in the framework of a standard `misaligned blazar' scenario.  We combined the new X-ray and $\gamma$-ray data presented above with archival radio, optical, and X-ray data from the NASA Extragalactic Database (NED), {\em XMM-Newton} optical monitor (OM) data from \citet{gli08}, and {\em Hubble Space Telescope} data from \citet{chi02}. The resulting SEDs are presented in Figure\,\ref{sed}. {It seems fairly clear, looking at the figure, that the nonthermal synchrotron peak in both targets lies between the core optical and X-ray emission, i.e., between $10^{15}$\ and $10^{17}$\,Hz, and we elaborate more on this point further below. }

We fit the \citet{gli08} and \citet{chi02} optical, {\em Suzaku} X-ray, and LAT $\gamma$-ray data {assuming they originate as non-thermal synchrotron/SSC from a relativistic jet} with the one-zone synchrotron/SSC model from \citet{fin08}. The resulting model curves are shown in Figure\,\ref{sed} and the model parameters are listed in Table\,\ref{sedfitpara}. See \citet{fin08} for a description of the model parameters and other model details.  {In our modeling we do not include a component from the disk/corona, consistent with our results from Section\,4.1. We also did not fit the radio data,} as this is likely to be from a superposition of self-absorbed jet components unrelated to the rest of the SED \citep{kon81}, and as such should be considered as upper limits in our one-zone synchrotron/SSC modeling. {The near infrared/integrated optical segments of the broadband spectra are clearly dominated by host galaxies, and therefore in our modeling we added a template of a giant elliptical from \citet{sil98}, adjusted to the redshifts of the analyzed sources; this template reproduces the optical data well.}  We assumed a relatively large jet angle to the line of sight ($\theta$ in Table\,\ref{sedfitpara}), consistent with the sources being misaligned BL Lacs, and used two variability timescales to test the robustness with respect to this poorly-constrained parameter.  {The models with two different variability timescales are given in Table \ref{sedfitpara}.}  Also listed in Table\,\ref{sedfitpara} are the results of one-zone synchrotron/SSC models applied to reproduce several other LAT-detected FR\,I radio galaxies from the literature. Model parameters for \pks\ and \tc\ are consistent with those for other radio galaxies which have been modeled previously, as shown in the table. The parameters $\Gamma$ and $\delta$ are lower than typically found in models of BL Lacs. We note that the black hole mass in \pks\ is estimated to be $10^{9.2} \, M_{\odot}$ \citep{bet03}, and in \tc\ as $10^{8.7} \, M_{\odot}$ \citep{rin05}; these are the typical values for radio galaxies \citep[$10^{8.1-9.5} \, M_{\odot}$;][]{mcl04} and BL Lac objects \citep[$10^{7.9-9.2} \, M_{\odot}$;][]{bar03}.

\begin{figure}[!t]
\begin{center}
\vspace{-1cm}
\includegraphics[scale=0.4]{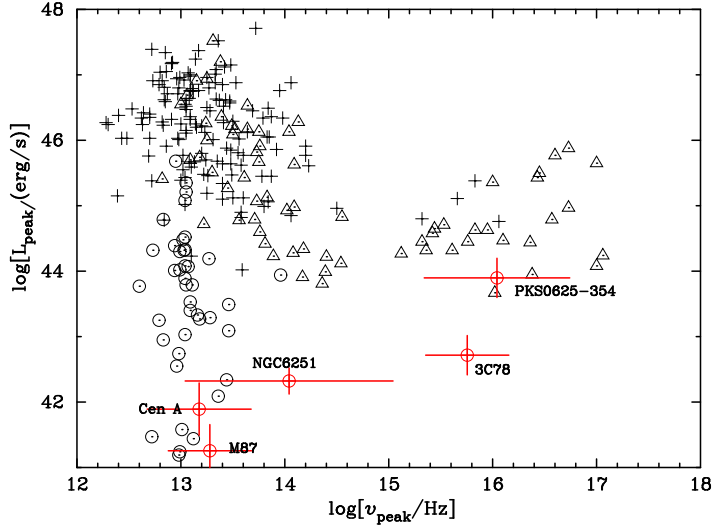}
\vspace{0.5cm}
\caption{Relation between synchrotron peak frequencies and peak luminosities of \pks\ and \tc, together with other sources from our sample of radio galaxies (red {circles}). For a comparison, radio galaxies, BL Lacs, and FSRQs from \citet{mey11} are also plotted (black circles, triangles, and crosses, respectively).}
\label{fplp}
\end{center}
\end{figure}

One major difference is the models for \pks, \tc, and NGC\,6251 \citep{mig11} have a higher $\gamma_{brk}$ by a factor of 10 compared to other radio galaxies in the table.  The larger $\gamma_{brk}$ leads to higher peak synchrotron frequencies and lower electron jet powers compared to magnetic jet powers. For Cen\,A, M\,87, and NGC\,1275, the models result in approximate equipartition between magnetic field and electron jet powers. The higher $\gamma_{brk}$ parameters for \pks\ and \tc\ are in turn mainly the result of the harder $\gamma$-ray spectra and soft X-ray spectra. Cen\,A, NGC\,6251, M\,87, and NGC\,1275 have soft LAT spectra ($\Gamma_{\rm HE}>2.1$), while the LAT spectra for \pks\ and \tc\ are harder ($\Gamma_{\rm HE}<2.1$), although NGC\,1275 is a borderline case. As we have already noted above, \pks\ and \tc\ were the hardest sources of the {\em Fermi}-LAT `misaligned AGN' list of \citet{abd10c}. The X-ray spectra for Cen\,A \citep{abd10a}, M\,87 \citep{abd09b}, NGC\,1275 \citep{abd09a}, and NGC\,6251 \citep{mig11} are hard ($\Gamma_{\rm X}<2$), indicating they originate from the SSC component, although M\,87 likely has some synchrotron contribution as well \citep{abd09b}. Assuming jet origin, the soft X-ray spectra ($\Gamma_{\rm X}>2$) for \pks\ and \tc\ indicates the X-rays originate from synchrotron emission, implying a high peak synchrotron frequency and $\gamma_{brk}$.

\citet{mey11} proposed a scenario where low-power jets (BL Lacs and FR\,I radio galaxies) have longitudinal bulk Lorentz factor gradients. In this scenario, when viewing more aligned jets one observes the faster portion of the outflow resulting in high synchrotron peaked sources, while for progressively more misaligned sources one sees progressively slower portions of the jet and progressively lower synchrotron peak frequencies. \citeauthor{mey11} also argued that such gradients are absent in high-power jets (FSRQs and FR\,II radio galaxies). In Figure\,\ref{fplp}, we plot the synchrotron peak luminosity ($L_{peak}$) versus the synchrotron peak frequency ($\nu_{peak}$) for the sample analyzed by \citeauthor{mey11} (see Figure\,4 therein), along with the results of model fits from the literature and from this paper for $\gamma$-ray bright radio galaxies. {Error bars on $\nu_{peak}$ and $L_{peak}$ were found from visual inspection of the SEDs.} We do not include here NGC\,1275, since its synchrotron peak is poorly constrained \citep{abd09a}. The sources \pks\ and \tc\ are found to have relatively high values of $\nu_{peak}$, not expected in the framework of the scenario of \citet{mey11}. This seems to disfavor to some extent their model, which states that high-peaked sources are only the most aligned jets.

It should be emphasized here that the values of $\nu_{peak}$ and $L_{peak}$ from \citet{mey11} were found from polynomial fits to radio, optical, and X-ray data, while values derived or adopted by us for $\gamma$-ray bright radio galaxies are found from a synchrotron/SSC model fit.  Our model fits are more physically-motivated, but also come with additional assumptions.  From the point of view of the Lorentz factor gradient scenario, our models are probably preferred, since this scenario assumes synchrotron/SSC emission. \tc\ is included in the sample of \citet{mey11}, but they obtained significantly lower values for $\nu_{peak}$ than we did (see their Table 3).  We believe this is for two reasons: (i) their phenomenological model fit the radio data, while our synchrotron models do not; and (ii) the inclusion of the hard LAT data spectra require high values of $\gamma_{brk}$, which result in high values for $\nu_{peak}$.  The latter indicates that LAT observations can be important for modeling the synchrotron portion of a radio loud AGN, even though the $\gamma$-rays are not directly produced by synchrotron emission.

Finally, we note there is some ambiguity as to whether \pks\ is a BL Lac object or a radio galaxy.  The optical spectrum of \pks\ resembles that of a BL Lac \citep{wil04}, although its radio morphology resembles an FR\,I radio galaxy \citep{ojh10}. \pks\ possesses a relatively bright unresolved core \citep{gov00}, as does \tc\ \citep{chi02}, and all the LAT-detected radio galaxies in Table\,\ref{sedfitpara}. They probably all have intermediate jet alignments, with $\theta$ in the range $\sim10\arcdeg$ to $30\arcdeg$.

\begin{table}[!t]
{\scriptsize
\begin{center}
\caption{SED model parameters of radio galaxies}
\label{sedfitpara}
\vspace{0.2cm}
\begin{tabular}{c|cc|cc|c|c|c|c} \hline
\hline
 & \multicolumn{2}{c}{PKS\,0625$-$354} \vline & \multicolumn{2}{c}{3C\,78} \vline & Cen\,A & M\,87 & NGC\,1275 & NGC\,6251 \\
\hline
$\Gamma$ & 5.8 & 5.7 & 2.93 & 5.75 & 7.0 & 2.3 & 1.8 & 2.4 \\
$\delta$ & 5.8 & 5.8 & 2.92 & 5.75 & 1.0 & 3.9 & 2.5 & 2.4 \\
$\theta$ [deg] & 10 & 19 & 20 & 20 & 30 & 10 & 25 & 25 \\
$B$ [G] & 0.82 & 0.11 & 0.77 & 0.02 & 6.2 & 0.055 & 0.05 & 0.04 \\
$t_v$ [Ms]  & $0.1$  & $1$  & $0.1$ & $1$ & $0.1$ & $1.2$ & $30$ & $1.7$ \\
$R_b$ [$10^{16}$\,cm] & $1.6$ & $16$ & $0.85$ & $17$ & $0.3$ & $1.4$ & $200$ & $12$ \\
\hline
$p_1$ & 2.5 & 2.5 & 2.7 & 2.7 & 1.8 & 1.6 & 2.1 & 2.75 \\
$p_2$ & 3.5 & 3.5 & 3.7 & 3.7 & 4.3 & 3.6 & 3.1 & 4.0 \\
$\gamma_{min}$ & $6\times10^3$ & $6\times10^3$ & $1\times10^3$ & $1\times10^4$ & $3\times10^2$ & 1 & $8\times10^2$ & 250 \\
$\gamma_{max}$ & $2\times10^6$ & $6\times10^6$ & $2\times10^7$ & $2\times10^7$ & $1\times10^8$ & $1\times10^7$ & $4\times10^5$ & $4.4\times10^5$ \\
$\gamma_{brk}$ & $2.9\times10^4$ & $4.6\times10^4$ & $7.3\times10^4$ & $1.4\times10^5$ & $8\times10^2$ & $4\times10^3$ & $9.6\times10^2$ & $2.0\times10^4$ \\
\hline
$P_{j,B}$ [$10^{42}$\,erg s$^{-1}$] & $43$ & $740$ & $0.3$ & $2.5$ & $65$ &  $0.02$ & $230$ & $0.4$ \\
$P_{j,e}$ [$10^{42}$\,erg s$^{-1}$] & $2$ & $10$ & $0.6$ & $13$ & $31$ & $7$ & $120$ & $160$ \\
\hline
\end{tabular}
\end{center}
The model parameters are as follows: $\Gamma$ is the bulk Lorentz factor, $\delta$ is the Doppler factor, $\theta$ is the jet angle, $B$ is the magnetic field, $t_v$ is the variability timescale, and $R_b$ is the comoving blob size scale, $p_1$ and $p_2$ are the low-energy and high-energy electron spectral indices, respectively, $\gamma_{min}$, $\gamma_{max}$, and $\gamma_{brk}$ are the minimum, maximum, and break electron Lorentz factors, respectively, and $P_{j,B}$ and $P_{j,e}$ are the jet powers in magnetic field and electrons, respectively. Rerefences: \citet[][for Cen\,A]{abd10a}, \citet[][for M\,87]{abd09a}, \citet[][for NGC\,1275]{abd10b}, \citet[][for NGC\,6251]{mig11}.
}
\end{table}

\section{Conclusions}

We have presented {\em Suzaku} results for nearby {\em Fermi}-LAT detected low-power radio galaxies, three of which are analyzed here for the first time.  Based on the Fe-K and X-ray spectral slope, X-ray variability, and [O III] line strength, we argued for the jet origin of the observed X-ray emission in \pks\ and \tc. {This conclusion is in agreement with the optical spectral classification of both AGN as ``low-excitation radio galaxies''.} We have modeled the broadband SEDs of these two objects including the most recent HE $\gamma$-ray spectra following from the analysis of the 5 year accumulation of the LAT data. We found that the bulk Lorentz factors of both sources are typical of those found from modeling the SEDs of FR\,I radio galaxies, and lower than typically found for BL Lac objects. The peak synchrotron frequencies for \pks\ and \tc\ are unusually high for radio galaxies, due to their unusually soft X-ray spectra and unusually hard LAT spectra. This seems at odds with the scenario outlined by \citet{mey11}, where high synchrotron peaked objects are the most aligned, and progressively less aligned objects have lower synchrotron peaks.  Further studies of \pks\ with very long baseline interferometry will help to clarify this issue (\citealt{mul13}, Truestedt et al.\ in preparation).

\acknowledgments

The authors thank {the anonymous referee for helpful comments that helped to improve the paper,} and the {\em Suzaku} and {\em Fermi} {teams} for the operation, calibration, and data processing. Y. ~F. was supported by JSPS KAKENHI Grant Numbers 2400000401 and 2424401400.  \L .~S. was supported by Polish NSC grant DEC-2012/04/A/ST9/00083.

The \textit{Fermi} LAT Collaboration acknowledges generous ongoing
support from a number of agencies and institutes that have supported
both the development and the operation of the LAT as well as
scientific data analysis.  These include the National Aeronautics and
Space Administration and the Department of Energy in the United
States, the Commissariat \`a l'Energie Atomique and the Centre
National de la Recherche Scientifique / Institut National de Physique
Nucl\'eaire et de Physique des Particules in France, the Agenzia
Spaziale Italiana and the Istituto Nazionale di Fisica Nucleare in
Italy, the Ministry of Education, Culture, Sports, Science and
Technology (MEXT), High Energy Accelerator Research Organization (KEK)
and Japan Aerospace Exploration Agency (JAXA) in Japan, and the
K.~A.~Wallenberg Foundation, the Swedish Research Council and the
Swedish National Space Board in Sweden.
 
Additional support for science analysis during the operations phase is
gratefully acknowledged from the Istituto Nazionale di Astrofisica in
Italy and the Centre National d'\'Etudes Spatiales in France.

This research has made use of the NASA/IPAC Extragalactic Database (NED) which is operated by the Jet Propulsion Laboratory, California Institute of Technology, under contract with the National Aeronautics and Space Administration.

\end{document}